\documentclass[aps,twocolumn,superscriptaddress]{revtex4}

\usepackage{graphicx}
\usepackage{dcolumn}
\usepackage{bm,gensymb,multirow}
\usepackage{lipsum}
\usepackage{natbib,amsmath}
\usepackage[normalem]{ulem}
\usepackage[T1]{fontenc}

\usepackage[dvipsnames]{xcolor}

\newcommand{\pwr}
{\affiliation{Department of Operations Research and Business Intelligence, Politechnika Wroclawska, Wybrzeze Stanislawa Wyspianskiego 27, 50370 Wroclaw, Lower Silesia, Poland}}

\newcommand{\cu}{\affiliation{Department of Physics, University of Calcutta, 92 Acharya Prafulla Chandra Road, Kolkata 700009, West Bengal, India}}

\begin{document}


\title{Persistence probability based dynamics and phase diagrams in biased q-voter models}


\author{Amit Pradhan}
\cu

\author{Pratik Mullick}
\pwr

\author{Parongama Sen}
\cu


\date{\today}

\begin{abstract}
Persistence probability in opinion dynamics models estimates the tendency of the agents  not to change their initial opinion till the present time. Here we consider two nonlinear $q$-voter models with binary opinions, where the dynamics are governed by a biased choice when the $q$ panel is not unanimous. The models are studied for different parameter ranges corresponding to the known stationary states. Mean field theory and numerical simulations are used to compute the persistence probability for the two types of opinion separately. The long time behavior in general is either a saturation or a decay that can be approximated by an exponential form, depending on the chosen parameters. Based on this, phase diagrams in the parameter space are presented for both the models. The regions in the phase diagrams indicate a strong correlation with the behavior of fixed points in the corresponding models, which is non-trivial as far as the persistence probability is concerned. 

\end{abstract}


\maketitle


\section{Introduction}\label{sec:intro}
Extensive research works have been made in the topic of persistence in
dynamical systems \cite{Satya_rev}.
Persistence is a phenomenon characterised by  the probability that a fluctuating non-equilibrium
field has not changed its sign up to  time $t$.
This phenomenon has been observed in various systems, for e.g., in magnetic systems
\cite{Derrida,Stauffer_1994}, simple diffusion
\cite{PhysRevLett.77.2867,HILHORST2000124,Brownian_bennaim},
fluctuating interfaces \cite{Surface_dasgupta,Surface_Krug,Surface_sakagawa} etc.

Originally, in the Ising-like spin models, where spins can take up binary values (typically $\pm 1$), and
the dynamical evolution involves spin flips at each time step,
 persistence probability $P(t)$ was defined as the probability that a spin has not changed its
 sign up to time $t$. In many cases a power law behavior of the persistence 
 probability is observed
and the associated exponent  is not related to any previously known static or
dynamic exponent. 



Applications of statistical physics tools have been found extensively in the field of sociophysics \cite{castellano2009statistical,sen2014sociophysics,jusup2022social}, with one of the most studied topics being opinion formation in a 
society.
In the opinion dynamics models, persistence can be defined as the probability that certain nodes or agents  have not changed their opinion since the beginning. The origin of time may correspond to a state where the opinions are randomly distributed. In real situations, it is normal to find persistent voters, and often those who change their opinion determine the swing in close contests.
Such a scenario is valid for e.g., the US presidential election \cite{biswas2017critical}.

Persistence probability is an important issue in opinion dynamics models and has been studied previously in quite a few models. Here, the variation is not necessarily a power law behavior,  for example in the Voter model \cite{voter_pers}
the behavior is very different, varying between a power law in one dimension and an exponential function in the mean field limit. In between, it decays faster than a power law but slower than an exponential function. 
In some other models, power law has been observed in one dimension \cite{Sznajd_pers,Soham2009} while for a kinetic exchange model in two dimensions, it shows an approximate power law at large times \cite{Sudip2020}. 

In this paper, we consider a fully connected agent based model for opinion dynamics in which a panel of $q$ agents is randomly picked up to update the opinion of the focal agent. Such a model was introduced in \cite{castellano2009nonlinear}. The update rule primarily depends on whether the $q$ panel is unanimous or not. A number of variations of the original model have been proposed henceforth. A  generalised model \cite{gen_qvoter}, DMSS model henceforth, from which several earlier versions can be recovered, was proposed recently.
Another model, which we call the MS model in this paper,  was studied in \cite{mullick2025social} where the composition of the panel, i.e., number of positive and negative opinions, was used to determine the the opinion updates when the $q$-panel is not unanibous. 
The latter model was also studied with the presence of contrarians \cite{appmps2026}. The details are provided in the next section.


Unlike in other models of opinion dynamics, there has been no study of persistence in $q$ voter models to the best of our knowledge. 
In the present work, we have calculated persistence probability by simulating an agent based model using the Monte Carlo method and also by using the mean field theory. Since the model involves a fully connected network, mean field theory is expected to yield reliable results. The persistence probabilities corresponding to positive and negative opinions have been studied separately. 

In the models considered here,   the steady state behavior have been explored already \cite{gen_qvoter,mullick2025social} showing that the fixed points could be consensus states with all opinions either positive/negative as well as other states depending on the parameter values. Even when the consensus states are the fixed points, it is not obvious how the persistence probability 
for either of the opinions will behave with time. The dependence on the panel size is also an important issue.

In the next section we discuss the models and the quantities calculated, as well as the bias mechanism to adopt the positive or negative opinion in terms of appropriate parameters. The analytical results from the mean-field theory are presented in Section \ref{sec:mft}, whereas the results from Monte Carlo simulations are presented in Section \ref{sec:simulation}. In Section \ref{sec:phasediagram} we present the obtained the phase diagrams for persistence probabilities, and compare it with the corresponding behavior of the fixed points obtained earlier. Finally we make some concluding remarks in the last section.












\section{Model description and features studied}
\label{sec:model}
We consider a population of $N$ agents with binary opinions, either positive or negative, which can be taken as $S= \pm 1$, interacting on a fully connected network. The system evolves over discrete time steps, and at each step, a randomly selected agent $A$ interacts with a group of $q\geq 2$ other agents, called a $q$-panel. 
We have studied two models, namely the DMSS model \cite{gen_qvoter}  and the MS model \cite{mullick2025social}. 


\textit{DMSS model}: This model was introduced in \cite{gen_qvoter}, which is a generalised $q$-voter model which maps to a few previously studied models, including the original $q$-voter model \cite{castellano2009nonlinear}.
Here,  if the $q$-panel is unanimous, the agent follows the opinion of the panel with probability $1$, this behaviour is called \textit{conformity}. 
If the $q$-panel is non-unanimous, then the update rule depends on the present opinion of the focal or target agent $A$ in a biased manner. If $A$ is in state $S=-1$, its opinion will flip with probability $\varepsilon_\uparrow$, and with the complementary probability $1 - \varepsilon_\uparrow$, it will remain in the original state. Similarly, an agent with state $S=1$  will change to $S=-1$  with probability $\varepsilon_\downarrow$, and with complementary probability $1 - \varepsilon_\downarrow$, it will stay as $S=-1$. The special case $\varepsilon_\uparrow = \varepsilon_\downarrow$ corresponds to the original $q$-voter model  \cite{castellano2009nonlinear}.


\textit{MS model}: In this model, a unanimous $q$ panel is treated as in the generalized nonlinear $q$-voter model, with the focal agent adopting the unanimous opinion. For a non-unanimous $q$-panel, the two opinions carry different influential powers. Here the opinion adopted by the focal agent is determined by a weighted average of the opinions in the $q$ panel. Specifically, opinion $+1$ is assumed to have a weight factor $p$, while opinion $-1$ has a weight factor equal to $1-p$. If $n$ out of $q$ agents in the panel support $+1$, then the focal agent adopts the positive opinion with probability
\begin{equation}
\label{p_q+}
    p_{q+} = \frac{np}{np + (q-n)(1-p)},
\end{equation}
and the negative opinion with probability $1-p_{q+}$.

The parameters in the DMSS model are $q, \varepsilon_\uparrow$ and $\varepsilon_\downarrow$ and in the MS model  there is only one parameter $p$ apart from $q$. 
The stability of the fixed points, studied earlier
showed a  strong $q$ dependence in the DMSS model, while in the MS Model there is appreciable $q$ dependence only for small values of $q$.

The two models considered here have a common factor of bias.
The bias in the first model comes from the unequal probabilities 
of flipping the positive/negative opinions when the $q$ panel is non-unanimous, depending on the opinion of the target agent. Suppose the current fraction of agents with 
positive opinion is $c$. The total probability of the updated opinion to be positive and negative respectively when the $q$ panel is non-unanimous are therefore 
$c(1- \varepsilon_\downarrow)+(1- c)\varepsilon_\uparrow$ and  $(1-c)(1-\varepsilon_\uparrow) +c\varepsilon_\downarrow$, which are in general unequal. The bias in the second model is  given by $p$ which is independent of the 
current state of the focal agent.


Persistence is defined as the fraction of agents that have never changed their opinion  up to a given time. Since there are two possible opinions, $+1$ and $-1$, we define separate persistence probabilities for each. Let $N_+(t)~(N_-(t))$ out of the total number $N$ of agents denote the number of agents that started with opinion $+1~(-1)$ and have not changed their opinion up to time $t$. The corresponding persistence probabilities are then defined as 
\begin{equation}
    P_+(t) = \frac{N_+(t)}{N}, \quad P_-(t) = \frac{N_-(t)}{N}.
\end{equation}
 The total persistence probability is then given by $P(t) = P_+(t) + P_-(t)$.


Our objective is to calculate the persistence probability $P_+(t)$ and $P_-(t)$ separately for an initial configuration containing a fraction $c_0$ of positive  opinions, where $c_0 = N_+(0)/N$ is the initial fraction of positive opinions.  

In the mean field theory, one can set up the rate equations for the persistence probabilities  in the $N \to \infty $ limit.
However, it is not possible to obtain the solutions in closed forms for general values of $q$. 
We  report the results for the two cases $q=2,3$ and the limiting case $q \to \infty$ for which  analytical results can be obtained. 
These results can be  compared to the Monte Carlo simulation results. For other values of $q$, we present the results obtained from the numerical simulations. 

\section{Results from the mean field theory}\label{sec:mft}



We consider the persistence dynamics in two representative extensions of the $q$ voter model in the mean field limit, introduced  in Sec. \ref{sec:model}, namely the DMSS and MS models. For these two models, the corresponding  flip rates and the mean field rate equation for the fraction of agents with positive opinion $c(t)$ are given in Refs. \cite{gen_qvoter}  and \cite{mullick2025social} respectively.


The persistence probabilities $P_+(t)$ and $P_-(t)$ can be expressed in terms of the transition rates in the following way:

\begin{equation}
 \label{master equation for persistence}
 \frac{dP_+(t)}{dt} = -\omega_{+\to-} P_+(t), \quad \frac{dP_-(t)}{dt} = - \omega_{-\to+}P_-(t),  
\end{equation}
with initial conditions $P_+(0) = c_0$ and $P_-(0) = 1-c_0$.
$\omega_{+\to-}$ and $\omega_{-\to+}$ denote the  flip rates given in terms of only one dynamic variable which we take as  $c$, the density of agents with positive opinion.

Whenever an explicit solution for $c(t)$ is available, $P_+(t)$ and $P_-(t)$ can be expressed directly as  functions of time $t$. In the cases where an explicit time dependent solution for $c(t)$ is not available, we rewrite the persistence equations in terms of the density $c$ using the chain rule,
\begin{equation}
    \frac{dP_{\pm}}{dt} = \frac{dc}{dt}\frac{dP_{\pm}}{dc} = \dot{c}(c)\frac{dP_{\pm}}{dc},
\end{equation}


the persistence rate equations (\ref{master equation for persistence}) can be rewritten as 
\begin{equation}
    \frac{dP_+}{dc} = -\frac{\omega_{+\to-}(c)}{\dot{c}(c)}P_+, \quad \frac{dP_-}{dc} = -\frac{\omega_{-\to +}(c)}{\dot{c}(c)}P_-.
\end{equation}
Note that $\dot c$ is a function of $c$ alone. 

Integrating from $c_0$ to $c(t)$ yields the general solutions
\begin{equation}
\label{general solution of persistence}
\begin{split}
    P_+(t) = c_0\exp\left[-\int_{c_0}^{c(t)}\frac{\omega_{+\to-}(c)}{\dot c(c)} dc\right],\\ \quad P_-(t)= (1-c_0)\exp\left[-\int_{c_0}^{c(t)}\frac{\omega_{-\to+}(c)}{\dot c(c)} dc\right].
\end{split}
\end{equation}
In the following subsections, we apply this general formalism of Eq. (\ref{general solution of persistence}) to analyze persistence for both the models. Although the integrands appearing in the expressions of $P_+$ and $P_-$ are rational functions of positive opinion density $c$ for any finite $q$, closed form expressions in terms of elementary functions can be obtained only for $q=2,q=3$ and in the limiting case $q\to\infty$. For $q\geq4$, the denominator of the integrand is a polynomial of degree higher than two, and the resulting integrals cannot, in general, be reduced to elementary functions and require elliptic integrals. Therefore, for both the  models we restrict our exact analytical treatment to the cases $q=2,~q=3$ and $q\to\infty$.





\subsection{Persistence for $q=2$}
Below, we discuss the closed form analytical expressions for the persistence probabilities $P_+(t)$ and $P_-(t)$ for both models discussed in Sec.~\ref{sec:model} for $q=2$.


\subsubsection{The DMSS model}


Let $\gamma^+$ and $\gamma^-$ denote the total probabilities, given in Ref. \cite{gen_qvoter}, that during a single update the density of positive opinions increases or decreases by $1/N$, respectively. Since an increase (decrease) in the density can only occur through the flipping of a negative (positive) agent, the per-agent flip rates satisfy $\gamma^+=(1-c)\omega_{- \to +}$ and $\gamma^- = c\omega_{+ \to -}$. Thus, for $q=2$, the per-agent flip rates are

\begin{equation}
\begin{split}
\omega_{+\to-}(c) = (1-c)^2+2\epsilon_{\downarrow}c(1-c),\\ \omega_{-\to+}(c) = c^2+2\epsilon_{\uparrow}c(1-c),
\end{split}
\end{equation}
and the corresponding mean field rate equation for $c(t)$ \cite{gen_qvoter} is
\begin{equation}
\label{rate equation for model A}
\frac{dc(t)}{dt}=c(1-c)(ac+b),
\end{equation}
with  $a=2(1-\epsilon_{\uparrow}-\epsilon_{\downarrow})$ and $b=(2\epsilon_{\uparrow}-1)$.

This equation is not invariant under any non-trivial symmetry transformation, including $c\to 1-c$, since such a transformation changes the coefficients of the equation. However, under the combined transformation $c\to 1-c$ together with  the interchange of $\epsilon_{\uparrow}$ and $\epsilon_{\downarrow}$, the equation preserves its functional form, and is therefore form invariant (covariant).

As discussed in Ref. ~\cite{gen_qvoter}, the rate equation (\ref{rate equation for model A}) admits three possible fixed points. The two absorbing states $c^*=0$ and $c^*=1$, corresponding to complete consensus with negative or positive states, respectively, always exist. In addition, an interior fixed point
\begin{equation}
    c^* = -\frac{b}{a} = \frac{1-2\epsilon_{\uparrow}}{2(1-\epsilon_{\uparrow}-\epsilon_{\downarrow})}
\end{equation}
may exist, provided it lies in the physical interval $0<c^*<1$. This condition is satisfied whenever the numerator and denominator have the same sign, namely either 

(i) $\epsilon_{\uparrow}<1/2$ together with $\epsilon_{\uparrow}+\epsilon_{\downarrow}<1$, or 

(ii) $\epsilon_{\uparrow}>1/2$ together with $\epsilon_{\uparrow}+\epsilon_{\downarrow}>1$.

A linear stability analysis further shows that the absorbing fixed point $c^*=0$ is stable for $\epsilon_{\uparrow}<1/2$, while $c^*=1$ is stable for $\epsilon_{\downarrow}<1/2$. The stability of the interior fixed point is controlled by the coefficient $a$, it is unstable for $\epsilon_{\uparrow}+\epsilon_{\downarrow}<1$ and becomes stable only when $\epsilon_{\uparrow}+\epsilon_{\downarrow}>1$. Consequently, when both biases satisfy $\epsilon_{\uparrow}<1/2$ and $\epsilon_{\downarrow}<1/2$, the two consensus states are simultaneously stable, leading to bistability. In this regime, the interior fixed point exists but is unstable and therefore acts as a separatrix dividing the phase space into two basins of attraction. The final steady state is then selected by the initial condition, depending on which side of the separatrix the system is 
prepared initially.

\begin{figure}
    \includegraphics[width=\linewidth]{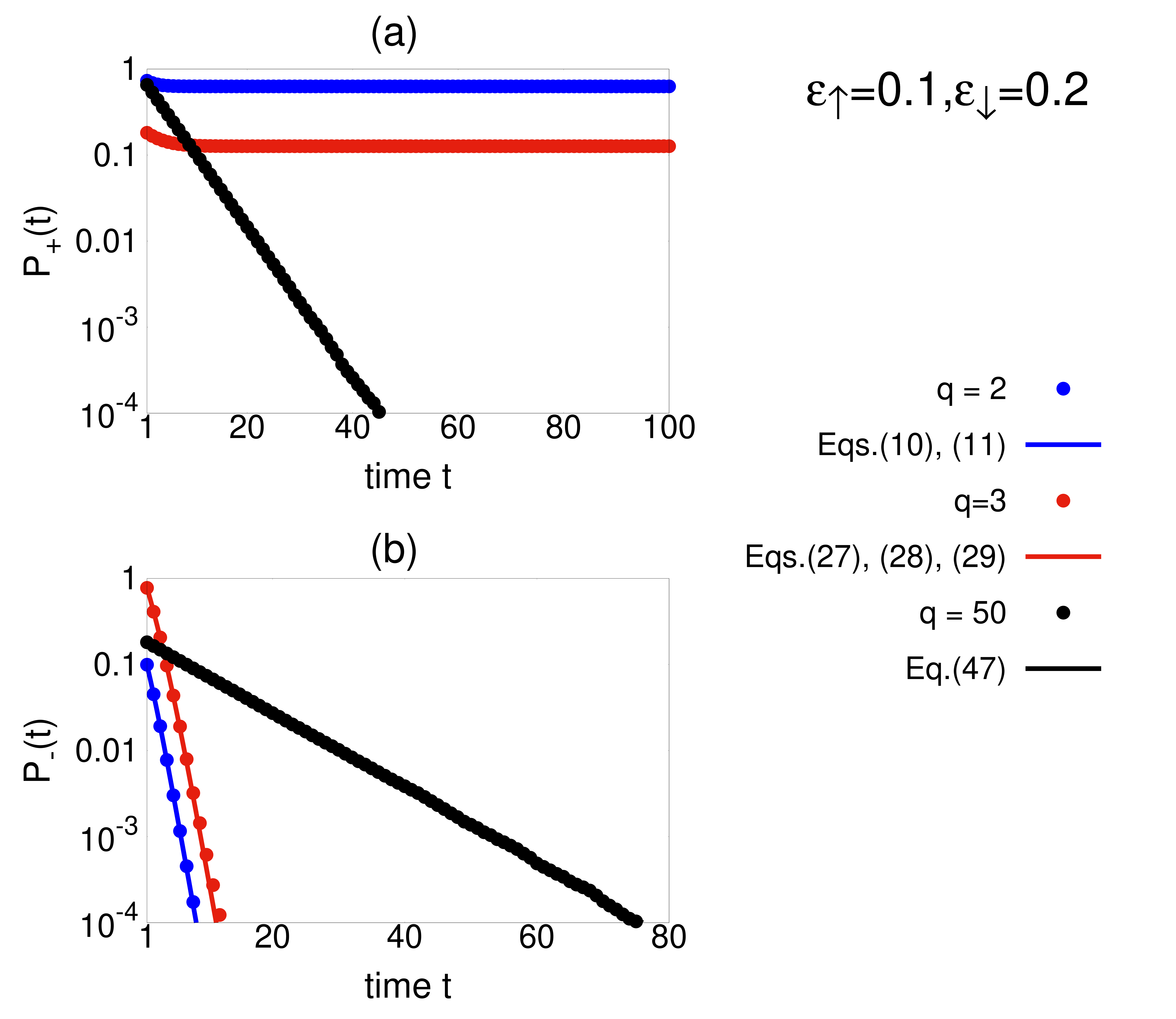}
    \caption{Time evolution of the persistence probabilities (a) $P_+(t)$ and (b) $P_-(t)$ for $q=2,3$ and $q=50$ for the DMSS model [see Sec. \ref{sec:model}]. Solid circles denote simulation data, while solid lines represent analytical calculations from the mean field theory: Eqs. (\ref{implicit solution of model A for q=2}) and (\ref{persistence_modelA_q=2}) for $q=2$, Eqs. (\ref{implicit solution of model A for q=3}), (\ref{P+ for q = 3}) and (\ref{P- for q = 3}) for $q=3$ and Eq. (\ref{persistence expression for modelA for q = 50}) in the $q\to \infty$ limit, which is well approximated by $q=50$ in simulations. Results are shown for bias parameters $\epsilon_{\uparrow}=0.1$ and $\epsilon_{\downarrow}=0.2$ and initial condition $c_0 = 0.8$. }
    \label{fig:pers_epsp0.1_epsm0.2_modelA}
\end{figure}

\begin{figure}
    \includegraphics[width=\linewidth]{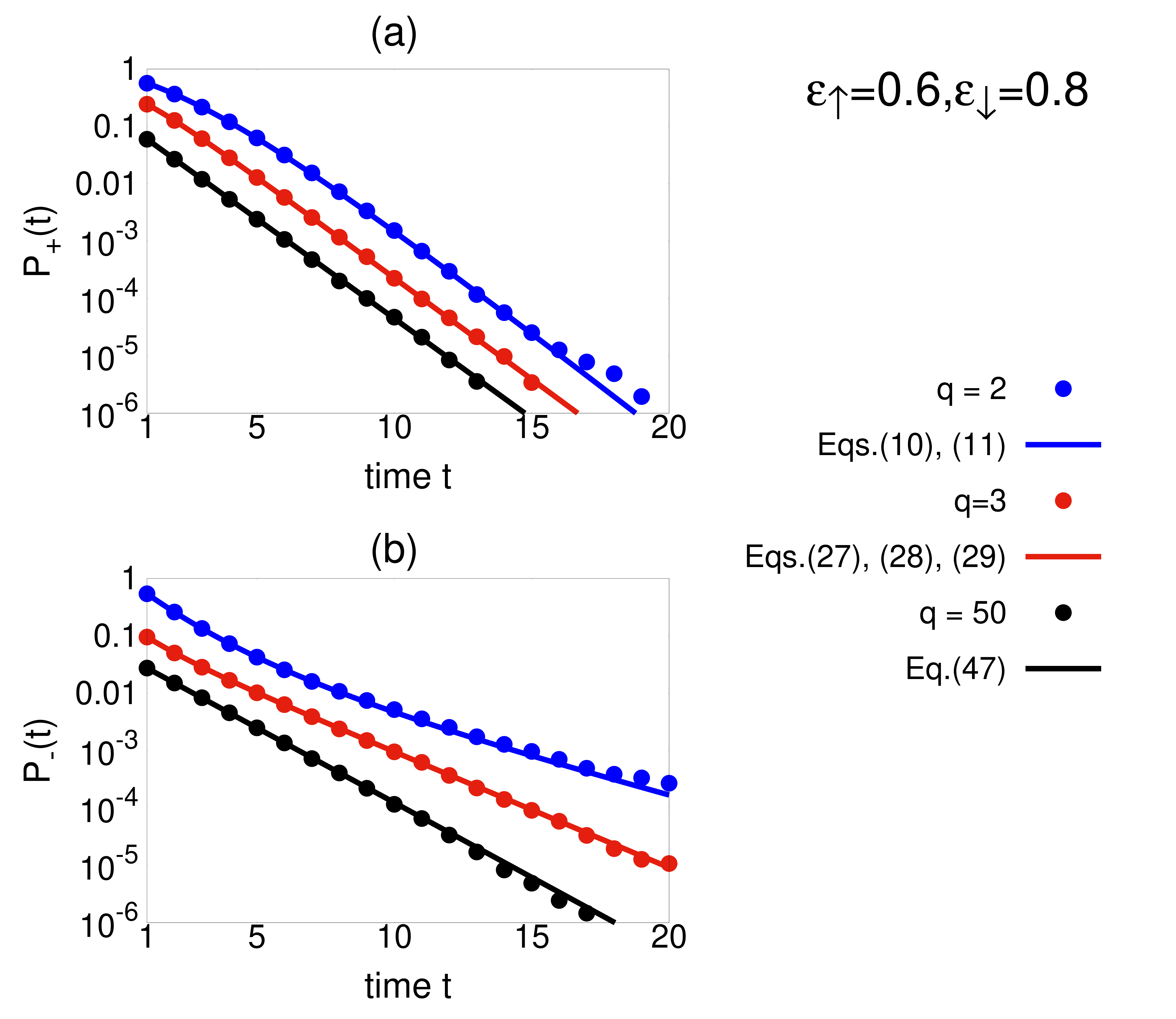}
    \caption{Time evolution of the persistence probabilities (a) $P_+(t)$ and (b) $P_-(t)$ for $q=2,3$ and $q=50$ for the DMSS model [see Sec. \ref{sec:model}]. Solid circles denote simulation data, while solid lines represent analytical calculations from the mean field theory: Eqs. (\ref{implicit solution of model A for q=2}) and (\ref{persistence_modelA_q=2}) for $q=2$, Eqs. (\ref{implicit solution of model A for q=3}), (\ref{P+ for q = 3}) and (\ref{P- for q = 3}) for $q=3$ and Eq. (\ref{persistence expression for modelA for q = 50}) in the $q\to \infty$ limit, which is well approximated by $q=50$ in simulations. Results are shown for bias parameters $\epsilon_{\uparrow}=0.6$ and $\epsilon_{\downarrow}=0.8$ and initial condition $c_0 = 0.8$. }
    \label{fig:pers_epsp0.6_epsm0.8_modelA}
\end{figure}

The implicit solution of Eq. (\ref{rate equation for model A}) obtained by separation of variables with initial condition $c(0)=c_0$ is
\begin{equation}
\label{implicit solution of model A for q=2}
    \frac{c^{a+b}}{(1-c)^b(ac+b)^a} = \frac{c_0^{a+b}}{(1-c_0)^b(ac_0+b)^a} 
    \exp\{b(a+b)t\}.
\end{equation}
Evaluating the integral given in Eq. \ref{general solution of persistence}, one obtains
\begin{equation}
\label{persistence_modelA_q=2}
\begin{split}
P_+(t) = c_0\left(\frac{c(t)}{c_0}\right)^{-\frac{1}{b}}\left(\frac{ac(t)+b}{ac_0+b}\right)^{\frac{1}{b}+\frac{a+b}{a}},\\ P_-(t) = (1-c_0)\left(\frac{1-c(t)}{1-c_0}\right)^{\frac{1}{a+b}}\left(\frac{ac(t)+b}{ac_0+b}\right)^{-\left(\frac{1}{a+b}+\frac{b}{a}\right)}.
\end{split}
\end{equation} The variations of $P_+(t)$ and $P_-(t)$ in the DMSS model for $q=2$ and for two combinations of $(\epsilon_{\uparrow},\epsilon_{\downarrow})$ are shown in Figures \ref{fig:pers_epsp0.1_epsm0.2_modelA} and \ref{fig:pers_epsp0.6_epsm0.8_modelA}.

Below we discuss the three important special cases:
\begin{itemize}
 \item 
(i) If we take $\epsilon_{\uparrow}+\epsilon_{\downarrow}=1$, in this case $a=0$ and the ODE (\ref{rate equation for model A}) reduces to the form
\begin{equation}
\label{rate equation for q = 2 in model A}
\frac{dc(t)}{dt} = bc(1-c) \quad \text{with} \quad b = 2\epsilon_{\uparrow}-1.
\end{equation}
The explicit solution is
\begin{equation}
  c(t) = \frac{1}{1+\frac{1-c_0}{c_0}\exp(-bt)}.  
\end{equation}


The corresponding persistence probabilities take the forms
\begin{equation}
\label{persistence expressions for model A for eps_up+eps_down = 1}
\begin{split}
 P_+(t) = c_0\left(\frac{c(t)}{c_0}\right)^{-\frac{1}{b}} \exp[c(t)-c_0]\quad \\ \quad P_-(t) = (1-c_0)\left(\frac{1-c(t)}{1-c_0}\right)^{\frac{1}{b}}\exp[-(c(t)-c_0)].
 \end{split}
\end{equation}


The special limit $\epsilon_{\uparrow}+\epsilon_{\downarrow}=1$ of the DMSS model for $q=2$ maps onto the MS model  for $q=2$ (with $p = \epsilon_\uparrow$), as will be discussed later in the context of the latter model.
For $b=0$~~($\epsilon_{\uparrow}=\epsilon_{\downarrow}=1/2$), these reduce smoothly to the mean field voter model results,  yielding the known exponential variations of the persistence  given in \cite{voter_pers} (note that a factor of $2$ is extra in the exponents in \cite{voter_pers}  which is probably an oversight). Exolicitly, for $b=0$, 


\begin{equation}
\label{voter result}
\begin{split}
 \lim_{b\to0}  P_+(t) = c_0\exp(-(1-c_0)t), \quad \\  \quad \lim_{b\to0} P_-(t) = (1-c_0)\exp(-c_0t).
 \end{split}
\end{equation}




\item
(ii) For $\epsilon_{\uparrow}=\epsilon_{\downarrow} = \epsilon$, this model reduces to the original $q$-voter model introduced by Castellano et al \cite{castellano2009nonlinear}; in this case $a=-2b$ with $b=2\epsilon-1$ and the mean field rate equation for $c$ is given in the implicit closed form

\begin{equation}
    \frac{c(1-c)}{(1-2c)^2} = \frac{c_0(1-c_0)}{(1-2c_0)^2}\exp(bt),
\end{equation}
(except for $c_0 \neq 1/2$).
Therefore, the closed form expressions for $P_+(t)$ and $P_-(t)$ depending on $c(t)$ take the forms
\begin{equation}
\begin{split}
 P_+(t) = c_0\left(\frac{c(t)}{c_0}\right)^{-\frac{1}{b}}\left(\frac{1-2c(t)}{1-2c_0}\right)^{\frac{b+2}{2b}}, \quad \\ \quad P_-(t) = (1-c_0)\left(\frac{1-c(t)}
 {1-c_0}\right)^{-\frac{1}{b}}\left(\frac{1-2c(t)}{1-2c_0}\right)^{\frac{b+2}{2b}} .
 \end{split}
\end{equation}
At the symmetric point $\epsilon=1/2$, the closed form expressions appear to be singular. However, the singularity is removable and taking the proper limit one recovers the mean field voter model result given in Eq. (\ref{voter result}).


\item
(iii) For $\epsilon_{\downarrow}=0$, this model is the one considered in 
\cite{muslim2024mass}.  In this case $a+b=1$ with $a=2(1-\epsilon_{\uparrow})$ and $b=2\epsilon_{\uparrow}-1$. The corresponding persistence probabilities are
\begin{equation}
\begin{split}
  P_+(t) = c_0\left(\frac{c(t)}{c_0}\right)^{-\frac{1}{b}} \left(\frac{ac(t)+b}{ac_0+b}\right)^{\frac{1}{ab}},\quad \\\quad P_-(t) = (1-c(t))\left(\frac{ac(t)+b}
  {ac_0+b}\right)^{-\frac{1}{a}},
  \end{split}
\end{equation}
where $c(t)$ can be found from the implicit solution given in Eq. (\ref{implicit solution of model A for q=2}) with $a+b=1$.
\end{itemize}


\subsubsection{The MS model }


For $q=2$, a selected panel is either unanimous or mixed. In the the latter case, the panel comprises one positive and one negative opinion. In a mixed panel the positive opinion is adopted with probability $p$ and the negative opinion is adopted with probability $1-p$. The flip rates are
\begin{equation}
\begin{split}
 \omega_{+\to-}(c) = (1-c)^2+2(1-p)c(1-c),\quad \\ \quad \omega_{-\to+}(c)=c^2+2pc(1-c),
 \end{split}
\end{equation}
As discussed in Ref.~\cite{mullick2025social}, the corresponding mean field rate equation for the concentration $c$ [see] Eq. (5) of Ref.~\cite{mullick2025social}] 
\begin{equation}
\label{rate equation for q = 2 in model B}
\frac{dc}{dt} = (2p-1)c(1-c),
\end{equation}
admits two fixed points $c^*=0$ and $c^*=1$, corresponding to negative and positive consensus, respectively, which exist for all $p\neq 1/2$. For $p<1/2$, $c^*=0$ is stable and $c^*=1$ unstable, while for $p>1/2$ the stability is reversed, with $c^*=1$ stable and $c^*=0$ unstable. Moreover, as reported in Ref.~\cite{mullick2025social}, this dynamical equation is invariant under the combined symmetry transformation $p\to 1-p$ and $c\to 1-c$, a property that holds for all values of $q$. For $p = 1/2$, the model reduces to the mean field voter model with $q=2$ neighbours. 

\begin{figure}
    \includegraphics[width=\linewidth]{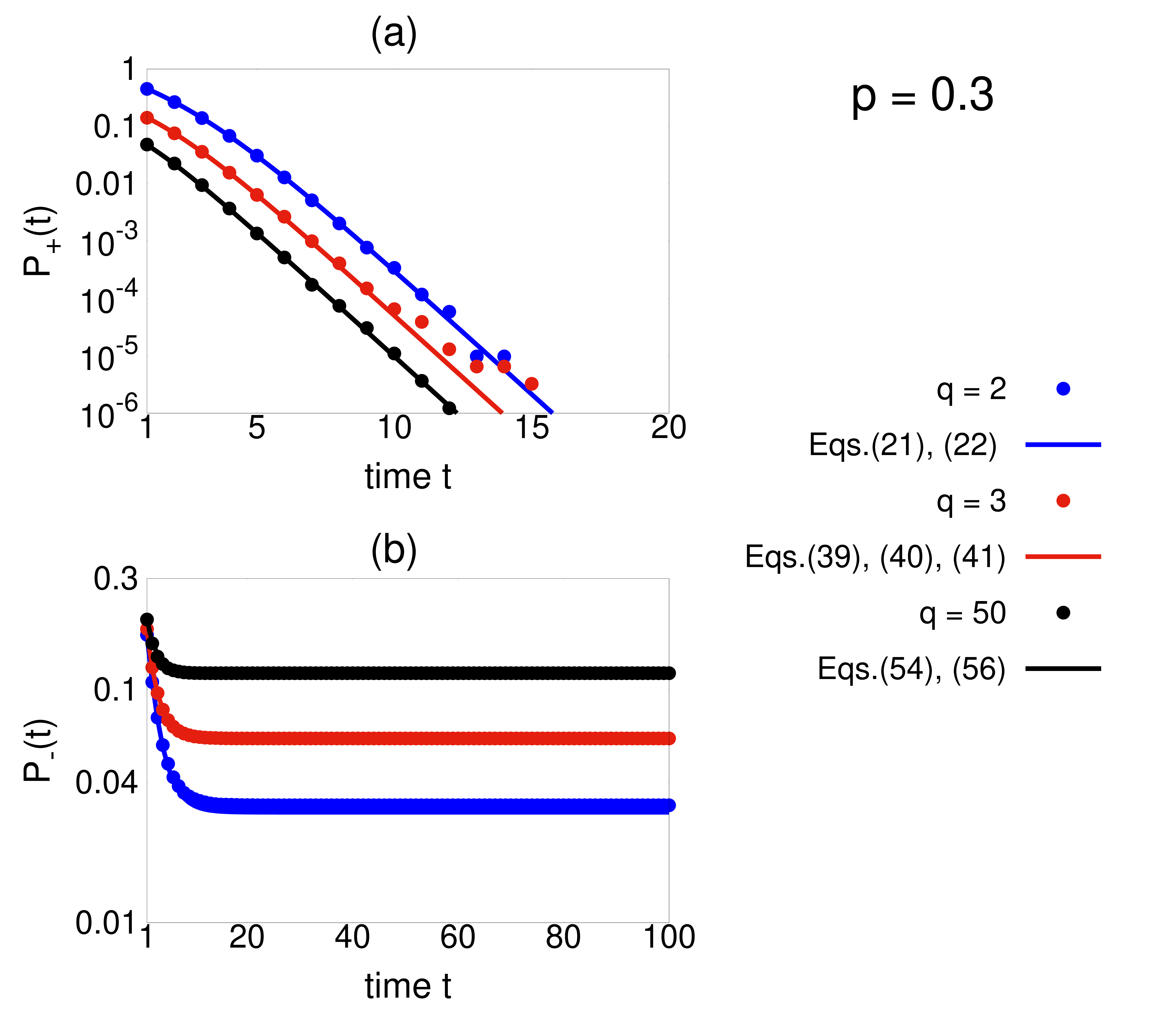}
    \caption{Time evolution of the persistence probabilities (a) $P_+(t)$ and (b) $P_-(t)$ for $q=2,3$ and $q=50$ for the MS model [see Sec. \ref{sec:model}]. Solid circles denote simulation data, while solid lines represent analytical calculations from the mean field theory: Eqs. (\ref{solution of rate equation for q = 2 for Mullick-Sen model}) and (\ref{persistence expression for model B.1 for q=2}) for $q=2$, Eqs. (\ref{implicit solution of model B for q=3}), (\ref{P+_q3_modelB}), (\ref{P-_q3_modelB}) for $q=3$ and Eqs. (\ref{implicit solution of c}) and (\ref{persistence expressions for modelB.1 for q=50}) in the $q\to \infty$ limit, which is well approximated by $q=50$ in simulations. Results are shown for bias parameter $p=0.3$ and initial condition $c_0 = 0.7$.}
    \label{fig:pers_p0.3_modelB}
\end{figure}

The ODE (\ref{rate equation for q = 2 in model B}) is separable and integrates to the explicit closed form

\begin{equation}
\label{solution of rate equation for q = 2 for Mullick-Sen model}
\begin{split}
\ln\left(\frac{c(t)}{1-c(t)}\right) = \ln\left(\frac{c_0}{1-c_0}\right)+(2p-1)t,\quad \\ \text{hence} \quad c(t) = \frac{1}{1+\frac{1-c_0}{c_0}\exp(-(2p-1)t)}.
\end{split}
\end{equation}
Using the flip rates and the rate equation and hence evaluating the integral given in Eq. (\ref{general solution of persistence}) gives the closed form of persistence probability for $p\neq 1/2$ as
\begin{equation}
\begin{split}
\label{persistence expression for model B.1 for q=2}
 P_+(t) = c_0\left(\frac{c(t)}{c_0}\right)^{-\frac{1}{2p-1}} \exp(c(t)-c_0),\quad \\ \quad P_-(t) = (1-c_0)\left(\frac{1-c(t)}{1-c_0}\right)^{\frac{1}{2p-1}}\exp(-(c(t)-c_0)). 
 \end{split}
\end{equation}
Here, the solution $c(t)$ can be found in Eq. (\ref{solution of rate equation for q = 2 for Mullick-Sen model}) in explicit form and substituting it yields $P_+,P_-$ as explicit functions of time. Owing to the symmetry of the dynamics, replacing $p\to1-p$ together with $c\to1-c$ interchanges $P_+(t)$ and $P_-(t)$, as expected from the exchange symmetry between positive and negative opinions. The variations of $P_+(t)$ and $P_-(t)$ in the MS model for $q=2$ and for a typical value of $p$ is shown in Figure \ref{fig:pers_p0.3_modelB}.

We note that, Eq. (\ref{rate equation for q = 2 in model B}) is identical to Eq. (\ref{rate equation for q = 2 in model A}) of  DMSS Model for $q=2$, provided the parameters satisfy $\epsilon_{\uparrow}+\epsilon_{\downarrow} = 1$ and the identification $\epsilon_{\uparrow}=p$ (and hence $\epsilon_{\downarrow} = 1-p$).


Under the same parameter mapping, the persistence probabilities in Eq. (\ref{persistence expression for model B.1 for q=2}) are also identical to Eq. (\ref{persistence expressions for model A for eps_up+eps_down = 1}) of  DMSS Model as expected. 

For $p=1/2$, one has $\frac{dc}{dt}=0$, so $c(t)=c_0$ for all $t$. The closed form expressions for the persistence appear to be singular at $p=1/2$ but this is a removable singularity. Taking the proper limit at $p=1/2$ gives the correct voter result
\begin{equation}
\begin{split}
 \lim_{p\to \frac{1}{2}}  P_+(t) = c_0\exp(-(1-c_0)t), \quad \\  \quad \lim_{p\to \frac{1}{2}} P_-(t) = (1-c_0)\exp(-c_0t),
 \end{split}
\end{equation}
which is the expected mean field voter behavior as this model reduces to voter model for $p=1/2$ for any $q$, noted earlier.

\subsection{Persistence for $\boldsymbol{q=3}$}
Below, we discuss the closed form analytical expressions for the persistence probabilities $P_+(t)$ and $P_-(t)$ for both models discussed in Sec.\ref{sec:model} for $q=3$. 


\subsubsection{The DMSS model}

Using the expressions for $\gamma^+$ and $\gamma^-$ for $q=3$ given in Ref. \cite{gen_qvoter}, the corresponding per-agent flip rates are
\begin{equation}
\label{flip rates for q = 3 in model A}
\begin{split}
    \omega_{+\to-}(c) = (1-c)^3 + 3\epsilon_{\downarrow}c(1-c), \quad \\ \quad 
    \omega_{-\to+}(c) = c^3 + 3\epsilon_{\uparrow}c(1-c).
\end{split}
\end{equation}
The corresponding mean field rate equation for the evolution of the density $c(t)$ is then given by 
\begin{equation}
\label{rate equation for q = 3 in model A}
    \frac{dc(t)}{dt} = c(1-c)(a^{\prime}+b^{\prime}c),
\end{equation}
with $a^\prime = 3\epsilon_{\uparrow}-1$ and $b^\prime = 2-3(\epsilon_{\uparrow}+\epsilon_{\downarrow})$.

Here also, as discussed in Ref.~\cite{gen_qvoter}, the rate equation (\ref{rate equation for q = 3 in model A}) admits three possible fixed points. The two absorbing states $c^*=0$ and $c^*=1$, corresponding to negative and positive consensus respectively, always exist. In addition, an interior fixed point
\begin{equation}
\label{interior_fixed_pt_q3_DMSS}
    c^* = -\frac{a^\prime}{b^\prime} = \frac{1-3\epsilon_{\uparrow}}{2-3(\epsilon_{\uparrow}+\epsilon_{\downarrow})}
\end{equation}
may exist, provided it lies in the physical interval $0<c^*<1$. This condition is satisfied whenever the numerator and denominator have the same sign, namely either 

(i) $\epsilon_{\uparrow}<1/3$ together with $\epsilon_{\uparrow}+\epsilon_{\downarrow}<2/3$, or 

(ii) $\epsilon_{\uparrow}>1/3$ together with $\epsilon_{\uparrow}+\epsilon_{\downarrow}>2/3$.

A linear stability analysis shows that the absorbing fixed point $c^*=0$ is stable for $\epsilon_{\uparrow}<1/3$, while $c^*=1$ is stable for $\epsilon_{\downarrow}<1/3$. The stability of the interior fixed point is controlled by the coefficient $b^\prime$, it is stable for $\epsilon_{\uparrow}+\epsilon_{\downarrow}>2/3$ and unstable otherwise. Consequently, when both biases satisfy $\epsilon_{\uparrow}<1/3$ and $\epsilon_{\downarrow}<1/3$, the two consensus states are simultaneously stable, leading to bistability. In this regime, the interior fixed point exists but is unstable and therefore acts as a separatrix dividing the phase space into two basin of attraction. The final steady state is then selected by the initial condition, depending on which side of the separatrix the system is prepared.

Imposing the initial condition $c(0)=c_0$, one obtains from Eq. (\ref{rate equation for q = 3 in model A})
\begin{equation}
\label{implicit solution of model A for q=3}
    \frac{c^{a^\prime+b^\prime}}{(1-c)^{a^\prime}(a^\prime+b^\prime c)^{b^\prime}} = \frac{c_0^{a^\prime+b^\prime}}{(1-c_0)^{a^\prime}(a^\prime+b^\prime c_0)^{b^\prime}} \exp(a^\prime(a^\prime+b^\prime)t).
\end{equation}

\begin{figure*}
    \centering
    \includegraphics[width=0.9\linewidth]{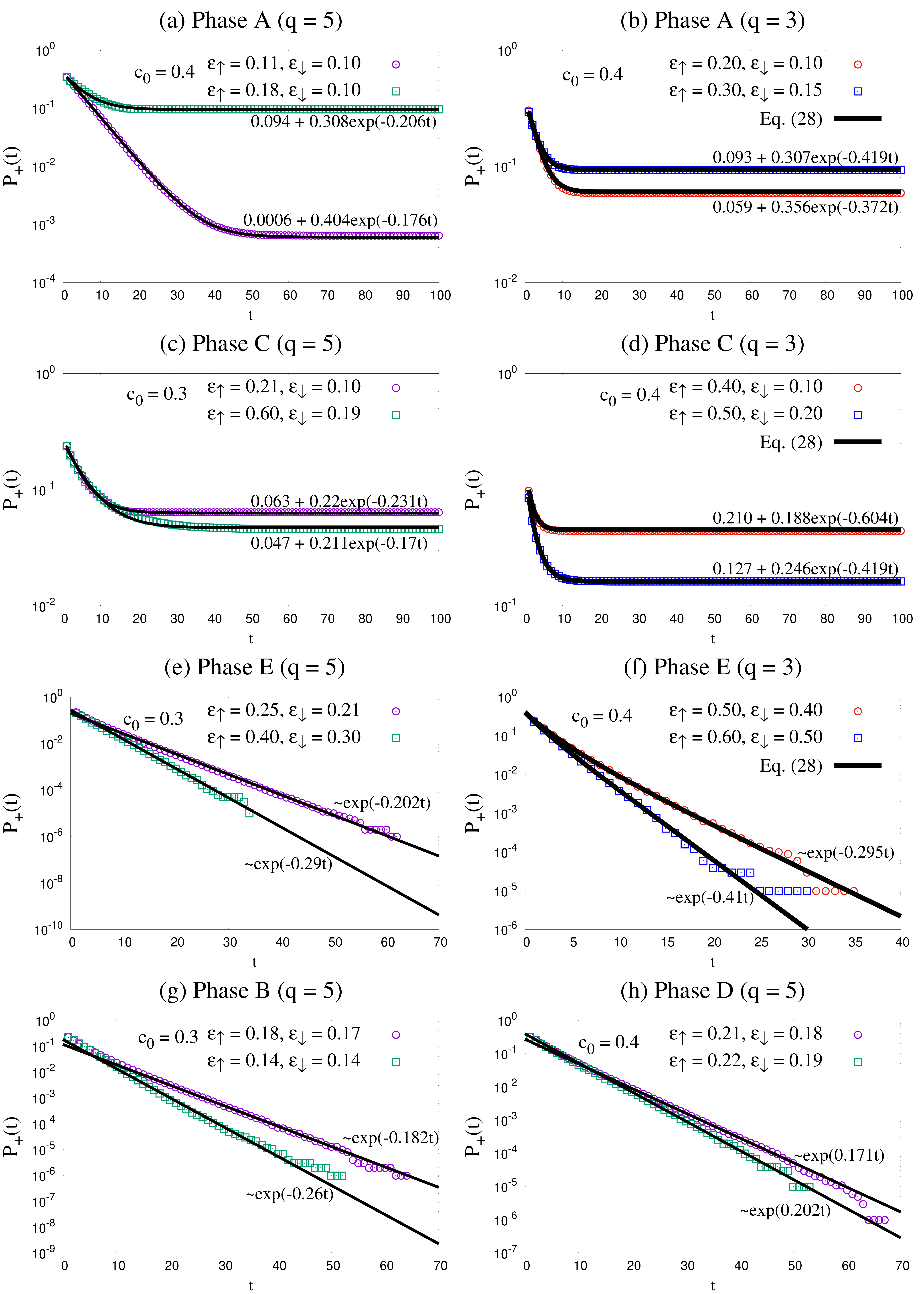}
    \caption{Persistence probability $P_+(t)$ is plotted as a function of time for the DMSS model in different phases of the phase diagram for $q=3$ and $q=5$. The phase nomenclature follows Ref. \cite{gen_qvoter}. The parameters $(\epsilon_{\uparrow},\epsilon_{\downarrow},c_0)$ are chosen such that the system reaches the positive consensus state in phases A and C, while it evolves to a non-consensus state in the remaining phases. Symbols denote Monte Carlo simulation results, and solid lines represents the analytical predictions whenever available. The persistence saturates in the consensus phases and decays exponentially in the non-consensus phases. }
    \label{fig:P_min}
\end{figure*}

Substituting the flip rates (\ref{flip rates for q = 3 in model A}) and rate equation for $c(t)$ (\ref{rate equation for q = 3 in model A}) in Eq. (\ref{general solution of persistence}), the integral can be evaluated exactly, yielding closed form expressions for persistence in terms of $c(t)$:
\begin{widetext}
\begin{equation}
\label{P+ for q = 3}
    P_+(c) = c_0\left(\frac{c(t)}{c_0}\right)^{-\frac{1}{a^\prime}}\left(\frac{a^\prime +b^\prime c(t)}{a^\prime + b^\prime c_0}\right)^{-\frac{1}{b^\prime}(3\epsilon_{\downarrow}-2-\frac{a^\prime}{b^\prime}-\frac{b^\prime}{a^\prime})}\exp\left[-\frac{(c(t)-c_0)}{b^\prime}\right],
\end{equation}
\begin{equation}
\label{P- for q = 3}
    P_-(c) = (1-c_0)\left(\frac{1-c(t)}{1-c_0}\right)^{\frac{1}{a^\prime+b^\prime}}\left(\frac{a^\prime +b^\prime c(t)}{a^\prime + b^\prime c_0}\right)^{-\frac{1}{b^\prime}(3\epsilon_{\uparrow}+\frac{a^\prime}{b^\prime}-\frac{a^\prime}{a^\prime+b^\prime})}\exp\left[\frac{(c(t)-c_0)}{b^\prime}\right].
\end{equation}
\end{widetext}
The variations of $P_+(t)$ and $P_-(t)$ in the DMSS model for $q=3$ and for two combinations of $(\epsilon_{\uparrow},\epsilon_{\downarrow})$ are shown in Figures \ref{fig:pers_epsp0.1_epsm0.2_modelA} and \ref{fig:pers_epsp0.6_epsm0.8_modelA}, and in Figure \ref{fig:P_min} we show the variation of $P_+(t)$ for several combinations of $(\epsilon_{\uparrow},\epsilon_{\downarrow})$.

Below we discuss the  important special cases:
\begin{itemize}
    \item 
    ~(i)For the symmetric case, $\epsilon_{\uparrow}=\epsilon_{\downarrow}=\epsilon$, this model reduces to the original Castellano model \cite{castellano2009nonlinear} and the parameters reduce to $a^\prime = 3\epsilon-1, b^\prime = 2-6\epsilon  = -2a^\prime$ so that $a^\prime+b^\prime = -a^\prime$. In this case, the persistence probabilities can be written explicitly in terms of the density $c(t)$ as
   \begin{widetext}
   \begin{equation}
    P_+(c) = c_0\left(\frac{c(t)}{c_0}\right)^{-\frac{1}{a^\prime}}\left(\frac{a^\prime +b^\prime c(t)}{a^\prime + b^\prime c_0}\right)^{\frac{1}{2}+\frac{3}{4a^\prime}}\exp\left[\frac{(c(t)-c_0)}{2a^\prime}\right],
\end{equation}
\begin{equation}
    P_-(c) = (1-c_0)\left(\frac{1-c(t)}{1-c_0}\right)^{-\frac{1}{a^\prime}}\left(\frac{a^\prime +b^\prime c(t)}{a^\prime + b^\prime c_0}\right)^{\frac{1}{2}+\frac{3}{4a^\prime}}\exp\left[-\frac{(c(t)-c_0)}{2a^\prime}\right].
\end{equation}
\end{widetext}
The time dependence of the density $c(t)$ follows from the implicit solution given in Eq. (\ref{implicit solution of model A for q=3}) by imposing the condition $a^\prime+b^\prime = -a^\prime$.

\item 
(ii) For the asymmetric case, $\epsilon_{\downarrow}=0$ and $\epsilon_{\uparrow} = \epsilon$, this model is identical to that considered in \cite{muslim2024mass} and the parameters reduce to $a^\prime = 3\epsilon-1,b^\prime = 2-3\epsilon$ so that $a^\prime + b^\prime =1$. Therefore, the persistence probabilities 
$P_+$ and $P_-$ take the forms
\begin{widetext}
\begin{equation}
    P_+(c) = c_0\left(\frac{c(t)}{c_0}\right)^{-\frac{1}{a^\prime}}\left(\frac{a^\prime +b^\prime c(t)}{a^\prime + b^\prime c_0}\right)^{\frac{1}{b^\prime}(2+\frac{a^\prime}{b^\prime}+\frac{b^\prime}{a^\prime})}\exp\left[-\frac{(c(t)-c_0)}{b^\prime}\right],
\end{equation}
\begin{equation}
    P_-(c) = (1-c_0)\left(\frac{1-c(t)}{1-c_0}\right)\left(\frac{a^\prime +b^\prime c(t)}{a^\prime + b^\prime c_0}\right)^{-\frac{1}{b^\prime}(3\epsilon+\frac{a^\prime}{b^\prime}-{a^\prime})}\\\exp\left[\frac{(c(t)-c_0)}{b^\prime}\right],
\end{equation}
\end{widetext}
where $c(t)$ can be found from the implicit solution given in Eq. (\ref{implicit solution of model A for q=3}) with $a^\prime+b^\prime = 1$.
\end{itemize}

\subsubsection{The MS model}


Setting $q=3$ in  the general expressions for the transition rates for arbitrary $q$ given in Ref. \cite{mullick2025social}, the flip rates can be written as
\begin{equation}
    \omega_{+\to -}(c) = (1-c)N_+(c), \quad \omega_{- \to +}(c) = cN_-(c),
\end{equation}
where
\begin{equation}
    N_+(c) = 1+a_1c+a_2c^2, \quad N_-(c) = b_0+b_1c+b_2c^2,
\end{equation}
with
\begin{equation}
\begin{split}
  a_1 = \frac{6(1-p)}{2-p} -2, \quad a_2 = 1-\frac{6(1-p)}{2-p}+\frac{3(1-p)}{1+p},\\ \quad b_0=\frac{3p}{2-p}, \quad b_1 = \frac{6p}{1+p}-\frac{6p}{2-p},\\\quad b_2 = 1+\frac{3p}{2-p}-\frac{6p}{1+p}.
  \end{split}
\end{equation}
As discussed in Ref. \cite{mullick2025social}, the corresponding mean field rate equation for the concentration $c$ is
\begin{equation}
\label{rate equation for q = 3 in model B}
    \frac{dc}{dt} = c(1-c)(A+Bc),
\end{equation}
where 
\begin{equation}
    A=\frac{3p}{2-p}-1, \quad B = 2-\frac{6}{2-p}+\frac{6p}{1+p}.
\end{equation}
Imposing the initial condition $c(0)=c_0$, one obtains from Eq. (\ref{rate equation for q = 3 in model B})
\begin{equation}
\label{implicit solution of model B for q=3}
    \frac{c^{A+B}}{(1-c)^{A}(A+Bc)^{B}} = \frac{c_0^{A+B}}{(1-c_0)^{A}(A+Bc_0)^{B}} \exp(A(A+B)t).
\end{equation}
Substituting the expressions for the flip rates and $\dot{c}$, and performing the integration in Eq. (\ref{general solution of persistence}), we obtain
\clearpage

\begin{widetext}
\begin{equation}
\label{P+_q3_modelB}
    P_+(t) = c_0\left(\frac{c(t)}{c_0}\right)^{-\frac{1}{A}}\left(\frac{A+Bc(t)}{A+Bc_0}\right)^{-\lambda}\exp\left[-\frac{a_2}{B}(c(t)-c_0)\right],
\end{equation}
and
\begin{equation}
\label{P-_q3_modelB}
  P_-(t) = (1-c_0)\left(\frac{1-c(t)}{1-c_0}\right)^{\frac{1}{A+B}}\left(\frac{A+Bc(t)}{A+Bc_0}\right)^{-\mu}\exp\left[\frac{b_2}{B}(c(t)-c_0)\right]. 
\end{equation}
\end{widetext}
The exponents appearing in the above equations are given by
\begin{equation}
    \lambda = \frac{1}{B}\left(a_1-\frac{B}{A}-\frac{A}{B}a_2\right), \quad \mu = \frac{1}{B}\left(b_0-\frac{A}{A+B}+\frac{A}{B}b_2\right).
\end{equation}
Here, $c(t)$ appearing in the above expressions is obtained from the implicit solution of Eq. (\ref{rate equation for q = 3 in model B}), given in Eq. (\ref{implicit solution of model B for q=3}). The variations of $P_+(t)$ and $P_-(t)$ in the MS model for $q=3$ and for a typical value of $p$ is shown in Figure \ref{fig:pers_p0.3_modelB}.

\subsection{Persistence for $\boldsymbol{q \to \infty}$}

Next, we discuss the closed form analytical expressions for the persistence probabilities $P_+(t)$ and $P_-(t)$ for both models discussed in Sec.\ref{sec:model} in the limit $q\to \infty$.

\subsubsection{The DMSS model}

From Ref. \cite{gen_qvoter}, one can obtain the transition probabilities governing the change in the density $c(t)$ of positive opinions for an arbitrary value of $q$. By normalizing these transition probabilities with respect to the fractions of positive and negative agents, the corresponding single agent flip rates for general $q$ are obtained as


\begin{equation}
    \omega_{+\to-}(c) =(1-c)^q+\epsilon_{\downarrow}[1-c^q-(1-c)^q],
\end{equation}
\begin{equation}
    \omega_{-\to+}(c)  = c^q + \epsilon_{\uparrow}[1-c^q-(1-c)^q].
\end{equation}


For any interior value $0<c<1$, the unanimous terms $c^q$ and $(1-c)^q$ vanish exponentially as $q\to \infty$, leading to constant flip rates $\omega_{+\to-} = \epsilon_{\downarrow}$ and $\omega_{-\to+} = \epsilon_{\uparrow}$.

The mean field evolution of $c(t)$ follows from
\begin{equation}
\label{rate equation for q tends to infinity in model A}
  \frac{dc(t)}{dt} = (1-c)\omega_{-\to+}-c\omega_{+\to-} = \epsilon_{\uparrow}-(\epsilon_{\uparrow}+\epsilon_{\downarrow})c. 
\end{equation}
Unlike the case of finite $q$, where absorbing consensus states at $c^*=0$ and $c^*=1$ exist, these fixed points disappear in the large $q$ limit. Instead, the dynamics admits a single stable fixed point at $c^*=\epsilon_{\uparrow}/(\epsilon_{\uparrow}+\epsilon_{\downarrow})$, corresponding to a mixed state. We emphasize that this $q\to \infty$ limit has not been reported or discussed in Ref.~\cite{gen_qvoter}, and is therefore analyzed here separately.

Eq. (\ref{rate equation for q tends to infinity in model A}) integrates to
\begin{equation}
    c(t) = c_{\infty} + (c_0-c_{\infty})\exp(-(\epsilon_{\uparrow}+\epsilon_{\downarrow})t),
\end{equation}
where $c_{\infty} = \frac{\epsilon_{\uparrow}}{\epsilon_{\uparrow}+\epsilon_{\downarrow}}$ is the stationary concentration of positive opinions.

The persistence probabilities therefore decay exponentially with time,
\begin{equation}
\label{persistence expression for modelA for q = 50}
    P_+(t) = c_0\exp(-\epsilon_{\downarrow}t); P_-(t) = (1-c_0)\exp(-\epsilon_{\uparrow}t).
\end{equation} The variations of $P_+(t)$ and $P_-(t)$ in the DMSS model for $q\rightarrow\infty$ and for two combinations of $(\epsilon_{\uparrow},\epsilon_{\downarrow})$ are shown in Figures \ref{fig:pers_epsp0.1_epsm0.2_modelA} and \ref{fig:pers_epsp0.6_epsm0.8_modelA}.

There are some special cases that can be discussed within this framework.
\begin{itemize}
\item
(i) When both flipping probabilities are equal, $\epsilon_{\uparrow}=\epsilon_{\downarrow}=\epsilon$ \cite{castellano2009nonlinear}, the concentration of positive agents relaxes exponentially towards the mixed state $c=1/2$ according to


\begin{equation}
    c(t) = \frac{1}{2} + (c_0-\frac{1}{2})\exp(-2\epsilon t).
\end{equation}
The corresponding persistence probabilities for positive and negative opinions also decay exponentially with the same rate $\epsilon$,
\begin{equation}
  P_+(t) = c_0\exp(-\epsilon t); P_-(t) = (1-c_0)\exp(-\epsilon t).   
\end{equation}
\item

(ii) On the other hand, when the bias acts only in favor of the positive opinion, i.e. $\epsilon_{\downarrow}=0$ and $\epsilon_{\uparrow}=\epsilon$ \cite{muslim2024mass}, the concentration of positive opinions increases monotonically as
\begin{equation}
    c(t) = 1-(1-c_0)\exp(-\epsilon t),
\end{equation}
leading to an exponential relaxation toward full positive consensus. The corresponding persistence probabilities then read
\begin{equation}
 P_+(t) = c_0; P_-(t) = (1-c_0)\exp(-\epsilon t).     
\end{equation}
\end{itemize}

\subsubsection{The MS model }


In the weighted influence $q$ voter model \cite{mullick2025social}, a mixed panel favors the positive opinion with relative weight $p$ and the negative opinion with weight $1-p$. In the mean field picture, the finite $q$ transition probabilities are binomial sums, but for large $q$ as $q\to \infty$, as $0<c<1$, the unanimous terms vanish, giving simple forms for the flip rates:
\begin{equation}
\begin{split}
 \omega_{-\to+}(c) = \frac{pc}{(1-p)(1-c)+pc}, \quad \\ \quad  \omega_{+\to-}(c) = \frac{(1-p)(1-c)}{(1-p)(1-c)+pc}.
 \end{split}
\end{equation}
The corresponding mean field equation [see Eq. (15) of Ref. \cite{mullick2025social}] for $c(t)$ is
\begin{equation}
    \frac{dc(t)}{dt} = c(1-c)\frac{2p-1}{(1-p)(1-c)+pc}.
\end{equation}
This differential equation is separable and integration yields
\begin{equation}
\label{implicit solution of c}
    \frac{c^{1-p}}{(1-c)^p} = \frac{c_0^{1-p}}{(1-c_0)^p}\exp((2p-1)t).
\end{equation}





\begin{figure*}
    \centering
    \includegraphics[width=0.9\linewidth]{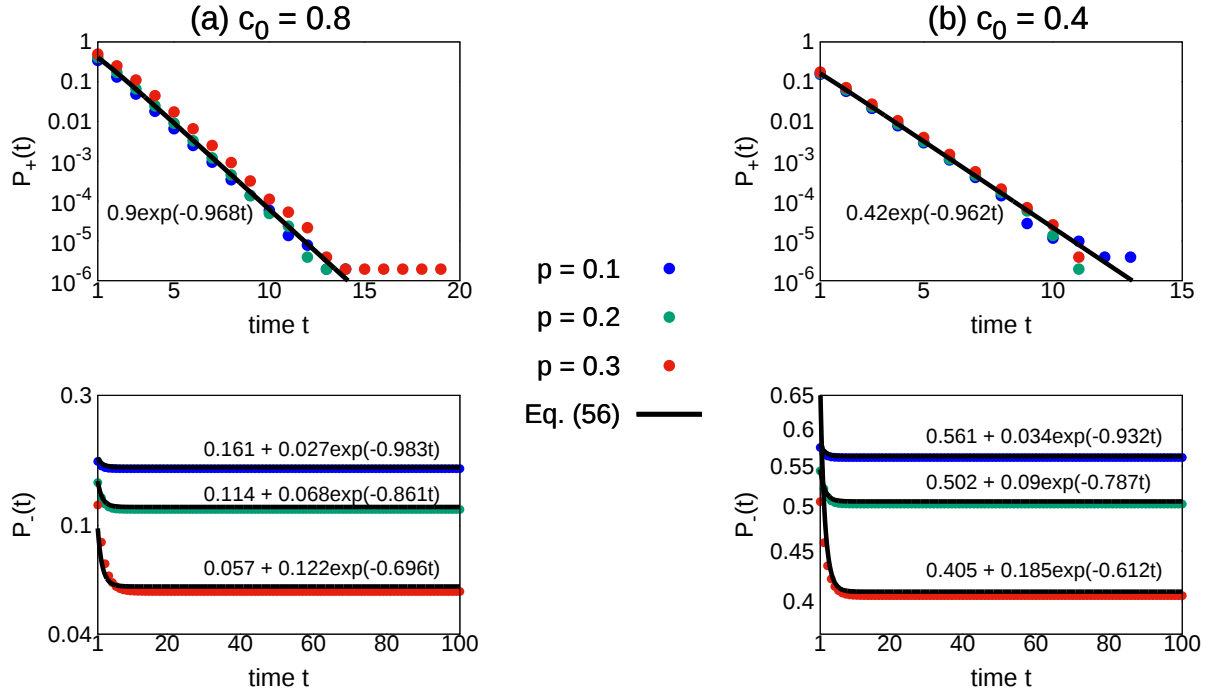}
    \caption{Persistence probabilities $P_+(t)$ and $P_-(t)$ for the MS model with $q=50$, shown on a log-linear scale. (a) $c_0 = 0.8$; (b) $c_0=0.4$. The results are shown for $p=0.1,0.2,$ and $0.3$. The simulation data are in excellent agreement with the analytical expressions given by Eq. (\ref{persistence expressions for modelB.1 for q=50}). For all values of $p$, the decay of $P_+(t)$ is well described by a universal exponential form, $P_+(t)\sim e^{-\lambda t}$, with nearly identical decay rates $\lambda \simeq 0.968$ for $c_0=0.8$ and $\lambda\simeq 0.962$ for $c_0=0.4$, indicating that the decay rate is essentially independent of both $p$ and the initial condition $c_0$. In contrast, $P_-(t)$ is well fitted by the form $\alpha+\beta e^{-\gamma t}$, with the fitting parameters depending on both $p$ and the initial condition $c_0$.} 
    \label{fig:MS_model_univ}
\end{figure*}

Substituting the expressions for $\omega_{+\to-},\omega_{-\to+}$ and $\dot c$, one obtains
\begin{equation}
\begin{split}
    \frac{\omega_{+\to-}(c)}{\dot c} = \frac{1-p}{(2p-1)c},\quad \\ \quad \frac{\omega_{-\to+}(c)}{\dot c} = \frac{p}{(2p-1)(1-c)}.
\end{split}
\end{equation}
Performing the above integration in Eq. (\ref{general solution of persistence}) gives,
\begin{equation}
\label{persistence expressions for modelB.1 for q=50}
\begin{split}
    P_+(t) = c_0\left(\frac{c(t)}{c_0}\right)^{-\frac{1-p}{2p-1}},\quad \\ \quad P_-(t) = (1-c_0)\left(\frac{1-c(t)}{1-c_0}\right)^{\frac{p}{2p-1}},
\end{split}
\end{equation}
where $c(t)$ can be found from the implicit closed from solution given in Eq. (\ref{implicit solution of c}). The variations of $P_+(t)$ and $P_-(t)$ in the MS model for $q\rightarrow\infty$ and for a typical value of $p$ is shown in Figure \ref{fig:pers_p0.3_modelB}, whereas in Figure \ref{fig:MS_model_univ} we show them for several values of $p$ and two typical values of $c_0$.

\section{Monte Carlo simulations}
\label{sec:simulation}

For $q=2,3$ we compared the analytical results with Monte Carlo Simulation, as shown in Figures \ref{fig:pers_epsp0.1_epsm0.2_modelA} to \ref{fig:pers_p0.3_modelB}. We see excellent agreement of the results between the two approaches. For $q \to \infty$, we compare the analytical results with those for a large $q$ value ($q=50$) in the simulation. 

For finite values of $q$, using numerical simulations is the only way
to obtain the results. However, for the MS model, there is no appreciable $q$ dependence, so we present results for the DMSS model mainly.

 Here we
recapitulate the phase terminology used in \cite{gen_qvoter} for the DMSS model: In phase A, 
 $c^* = 0,1$ are the only stable fixed points while in phases B, C \& D, stable fixed points at some intermediate value of 
$c$ also exist in addition to the consensus state. Phase E has no consensus states. 
The fixed points depend on the specific values of the parameters.

As an example, in Figure \ref{fig:P_min} we show some results for $q=3$ and $5$ for the DMSS model, where we focus on the behavior of the persistence probability for the initial minority. Our choice is $c_0 < 0.5$ in all the cases so we are 
presenting the results for $P_+(t)$ essentially. We choose $\varepsilon_\uparrow$ and $\varepsilon_\downarrow$ values 
corresponding to the phases mentioned above and the initial values of $c_0$ such that 
the final states are consensus states corresponding to $c=1$ in phases A and C for both $q=3$ and $5$. 
 In the  other  phases (E for $q=3$ and B, D, E for $q=5$), the final state is a non-consensus state for the choice of parameters (note that in phase E, it is always a non-consensus state which is stable). For $q=3$ we show results obtained from both Monte Carlo simulations and the exact results. 
 Typically, we find a saturation in phase A and C
while for the other phases, whenever the stable fixed point  is a non-consensus state, $P_+$ shows a decay in time which can be fitted accurately by
an  exponential form.



We obtain the significant result that although the qualitative behavior is similar, the quantitative dependence on time varies even within the same phase, i.e., the slopes of the exponential fits and the saturation values are dependent on the exact parameter values. 




 Similar to the DMSS model, the persistence dynamics in the MS model is governed by the nature of the steady state. As shown in Figure \ref{fig:MS_model_univ}, for $p<1/2$, the system evolves to the negative consensus state. Consequently, $P_+(t)$ decays exponentially to zero with a decay rate that is nearly independent of both $p$ and the initial density $c_0$. In contrast, $P_-(t)$ approaches a finite value of the form $\alpha+\beta e^{-\gamma t}$, 
with the fitting parameters depending strongly on both $p$ and $c_0$. For $p>1/2$, $P_+$ and $P_-$ will behave exactly in the opposite way. 
 Unlike the DMSS model, where the exponential decay rates vary continuously with the model parameters, the exponential decay associated with the minority persistence in the MS model is found to have an almost parameter independent decay rate over the range of parameters studied.


\section{Phase Diagrams}
\label{sec:phasediagram}

Based on the behavior of $P_{+}(t \to \infty)$  discussed in the previous section, it is possible to construct a phase diagram, where the phase boundary separates two regions: one in which $P_+(t\to\infty)$ attains a finite value, and the other in which it vanishes.


In general, the behavior of $P_+(t\to\infty)$ is found to be strongly correlated with the existence of a stable positive consensus state. $P_+(t\to \infty)$ saturates to a finite value whenever the system reaches the positive consensus state, while it decays to zero otherwise. The behavior of $P_-(t\to \infty)$ is complementary, and an analogous phase diagram can be obtained for it.

The persistence behavior also depends on the initial density in the DMSS model where it is known that for $\varepsilon_\uparrow, \varepsilon_\downarrow < \frac{1}{q}$, the system may reach 
 either of the consensus states, depending on the initial density.

Since we wish to correlate the behavior of $P_\pm (t \to \infty)$ with the stable consensus/non-consensus states, we present an alternative phase diagram for the two parameter DMSS model based on these states. In this representation, there are only three regions for any value of $q$: positive consensus, negative consensus, and an interior fixed-point phase. This differs from the representation  in \cite{gen_qvoter}, where the  phases were identified according to the number of stable and unstable fixed points, and in principle could be many in number.

 Below, we discuss specifically the cases for $q=3$ and 5 for the DMSS model. For $q=3$, one can obtain the phase 
 boundaries based on the saturation values of $P_+(t \to \infty)$  in closed form, while for $q=5$, these are estimated only by Monte Carlo simulations. 
 
 \begin{figure*}
    \centering
   \includegraphics[width=\linewidth]{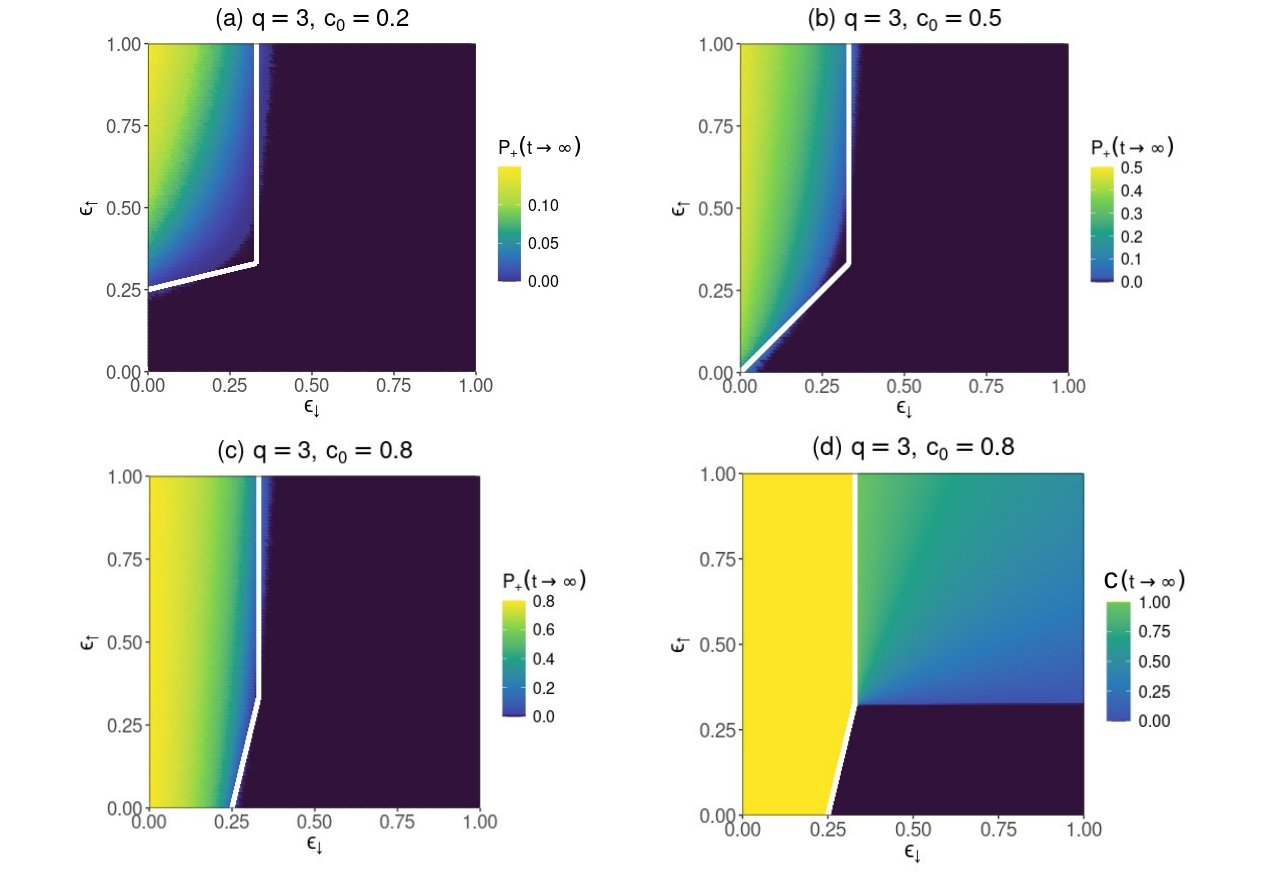}
    \caption{Heatmaps showing the variation of $P_+(t\rightarrow\infty)$ with $(\epsilon_\uparrow,\epsilon_\downarrow)$ for $q=3$ with different values of $c_0$ viz. (a) 0.2, (b) 0.5 and (c) 0.8. (d) Equilibrium phase diagram for regions with steady state values $c=1$, $c=0$ and $0<c<1$.}
    \label{fig:heatmap}
\end{figure*}

 For $q = 3$, when $\epsilon_{\uparrow},\epsilon_{\downarrow}<1/3$, the positive consensus state is reached for initial densities above the unstable fixed point, i.e., $c_0 > c^*$ [Eq. (\ref{interior_fixed_pt_q3_DMSS})], while for $c_0<c^*$ it evolves to the negative consensus state. We have verified numerically that $P_+(t\to\infty)$ remains finite for all initial conditions satisfying $c_0>c^*$, including values of $c_0$ very close to $c^*$. Hence the persistence phase boundary is determined by $c_0=c^*$. Using the expression for the unstable fixed point $c^*$ [Eq. (\ref{interior_fixed_pt_q3_DMSS})], the phase boundary is obtained as
 \begin{equation}
 \varepsilon_\uparrow = \frac{3c_0\epsilon_{\downarrow}+1-2c_0}{3-3c_0}.
 \label{boundary}
 \end{equation}
On the other hand, for $\varepsilon_\downarrow > 1/q$,  $P_{+}(t \to \infty) = 0 $ so that the phase boundary for 
 $\varepsilon_\downarrow > 1/q$ is just a vertical line at $1/q$. 
 We show the phase diagrams for $q=3$ and different values of $c_0$ 
 in Figure \ref{fig:heatmap}. For $q=3$, for $\varepsilon_\downarrow <  1/q$, the phase boundary is a straight line given by Eq. (\ref{boundary}) and dependent on $c_0$ 
 while for  $\varepsilon_\downarrow > 1/q$, it is independent of $c_0$.
 
We note that for larger values of $c_0$, the agreement with the theoretically obtained phase boundary and the one obtained in the present paper becomes better. This may be because at smaller values of $c_0$, the behavior at the points very close to the theoretical boundary becomes difficult to identify. A similar phase diagram, as mentioned above,  can be conceived for $P_{-}(t\to \infty)$. In addition to the phase diagrams obtained on the basis of the saturation values of $P_+(t \to \infty)$ shown in Figs \ref{fig:heatmap}(a)-(c) (together with the theoretical phase boundaries for the positive consensus region),  we  present the equilibrium phase diagram based on  positive/negative and non consensus regions for $c_0 = 0.8$ in Figure \ref{fig:heatmap}(d).  The latter is obtained by numerically solving the time dependent equation involving $c$. The negative consensus region in this diagram  can be identified as  the phase corresponding to $P_{-}(t\to \infty)$ having a non-zero value.

\begin{figure*}
    \centering
   \includegraphics[width=\linewidth]{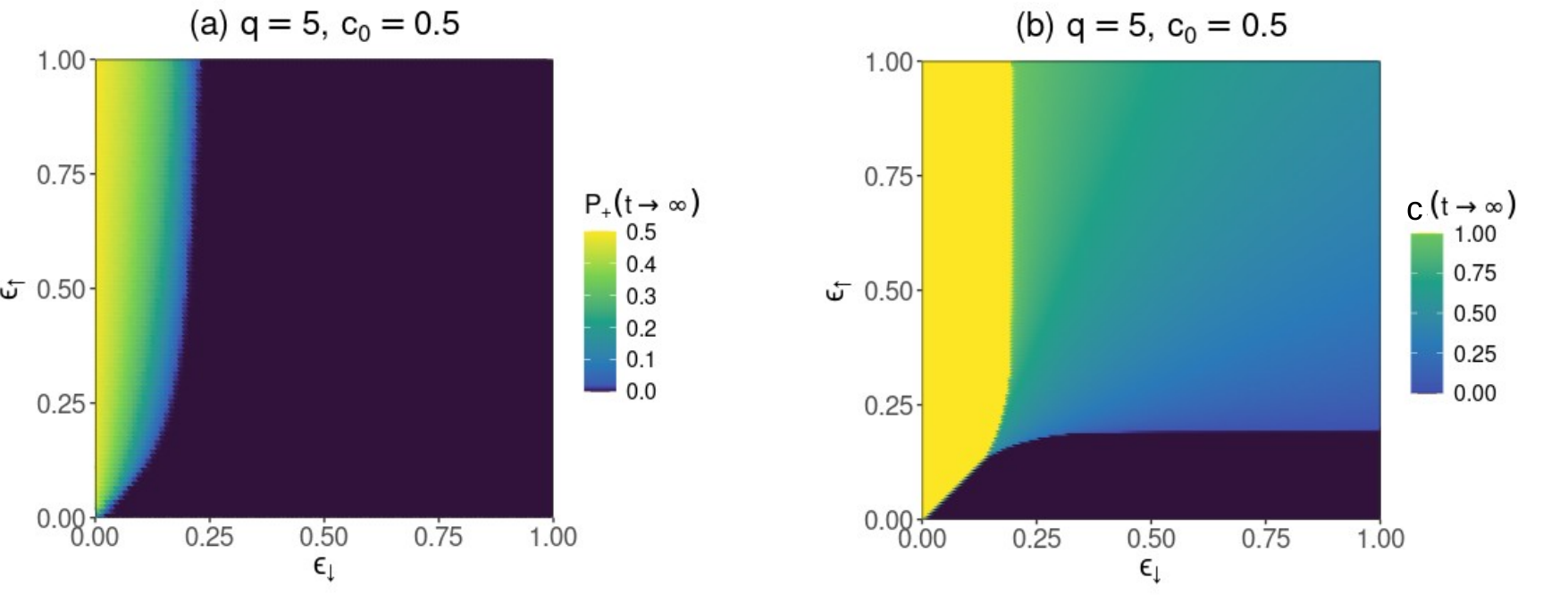}
    \caption{(a) A heatmap showing the variation of $P_+(t\rightarrow\infty)$ with $(\epsilon_\uparrow,\epsilon_\downarrow)$ for $c_0=1/2$ and  $q=5$.  (b) The equilibrium phase diagram shows regions with steady state values $c=1$, $c=0$ and $0<c<1$.}
    \label{fig:heatmap_q5}
\end{figure*}

 For $q=5$, it is not possible to use the same analysis as in $q=3$ to obtain the theoretical phase boundaries. Hence we show for comparison, the phase diagram showing the consensus and non-consensus regions obtained by solving numerically the dynamical equation for $c$. The results are presented in Fig \ref{fig:heatmap_q5}.

 For the MS model, it is also instructive to examine the asymptotic positive and negative persistence in the one parameter space by varying the influential parameter $p$. The corresponding analytical expressions for $P_+(t\to \infty)$ and $P_-(t\to \infty)$ can be obtained directly from Eqs. (\ref{P+_q3_modelB}) and (\ref{P-_q3_modelB}) by taking the large time limits $c(t\to \infty) = 1$ for $p>1/2$ and $c(t\to \infty)=0$ for $p<1/2$ respectively.

\begin{figure}
    \includegraphics[width=\linewidth]{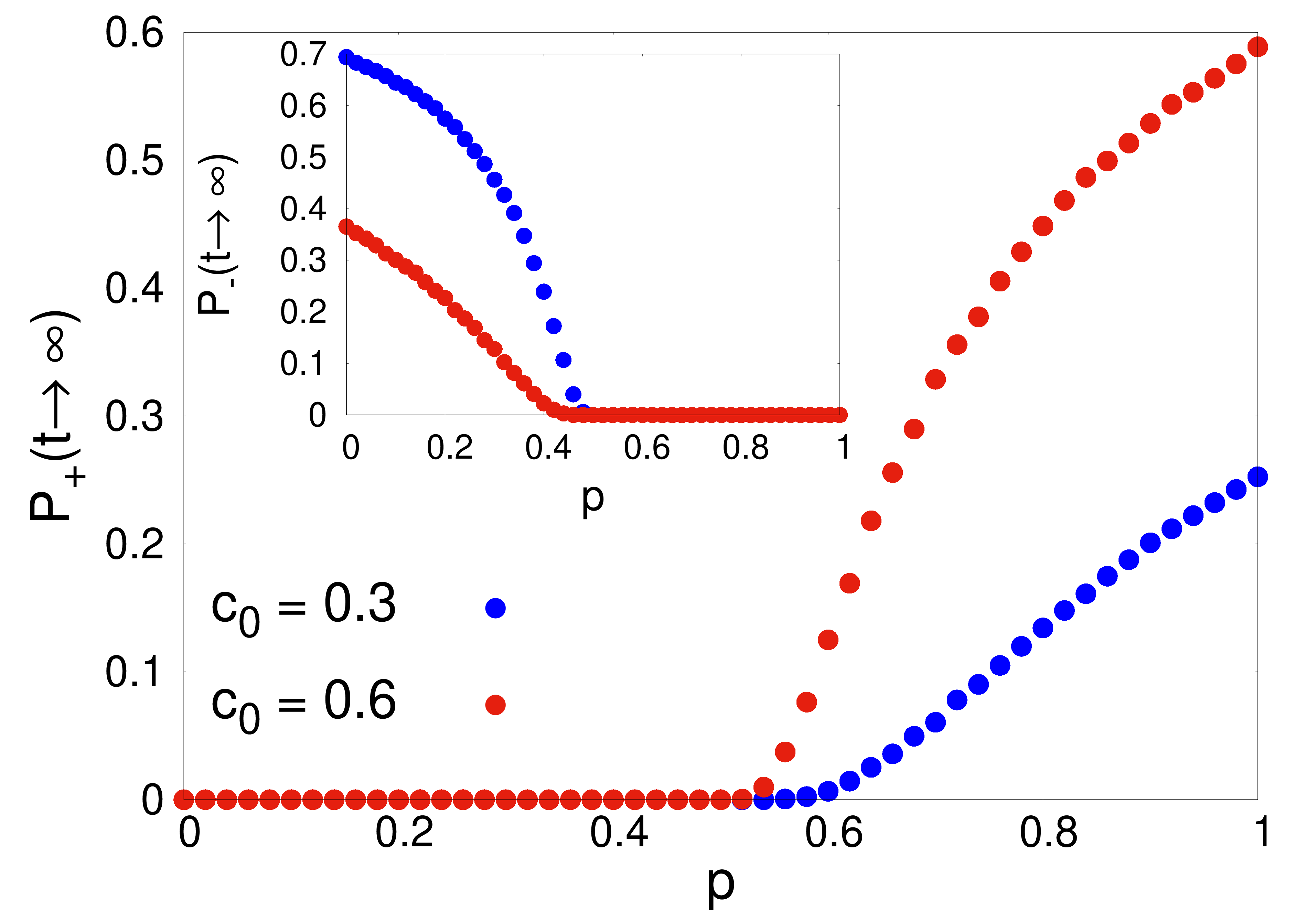}
    \caption{Asymptotic positive persistence probability $P_+(t\to \infty)$ is plotted as functions of the influence parameter $p$ for $q=3$ and two initial concentrations, $c_0=0.3$ and $0.6$, in the MS model. The inset shows the plot of the corresponding asymptotic negative persistence probability $P_-(t\to \infty)$ for the same initial concentrations.  }
    \label{phase_diag_MS_model}
\end{figure}

Figure \ref{phase_diag_MS_model} shows the variation of the asymptotic positive and negative persistence with $p$ for two representative initial concentrations, $c_0 = 0.3$ and $0.6$ for $q=3$. As expected, $P_+(t\to \infty)$ vanishes for $p<1/2$, while $P_-(t\to \infty)$ vanishes for $p>1/2$;  note that $p< 1/2$ ($p> 1/2$) region corresponds to the negative (positive) consensus region irrespective of the value of $c_0 \neq 0,1$.  The saturation values of both types of persistence probabilities increase monotonically in the relevant regions as the system is driven more strongly towards the corresponding consensus state, approaching finite values at $p=1$ and $p=0$, respectively. Another important observation is that although the qualitative behavior of the persistence probabilities is similar for any initial concentration $c_0$,  the asymptotic (non-zero) values  depend sensitively on it, as is evident from the two panels of Figure \ref{phase_diag_MS_model}. The asymptotic values,  e.g., for $P_+(t \to \infty)$ increases as $c_0$ increases
when $p> 0.5$. 
It is expected that the variations will be similar for any $q$ as the MS model results show negligible dependence on $q$. 

\section{Discussions and Conclusions}
\label{sec: conclu}

We note that the persistence behavior strongly depends on the stable fixed points in both MS and DMSS models.
To show this explicitly, we have presented  the equilibrium phase diagrams of the DMSS model in terms of consensus 
behavior, where there are  three regions for any value of $q$. This phase diagram is dependent on the initial value of $c$, the density of positive opinion.  On the other hand, for the MS model,
consensus states are simply determined by the value of $p$, irrespective of the initial concentration of
the two types of opinion. 

For regions in the phase diagram where a consensus state can be reached, 
typically  $P_+$  (for consensus with all agents adopting the positive opinion) 
or $P_-$ (for consensus with all agents adopting the negative opinion) reaches a saturation value 
very fast.

Mean field theory indicates some non-typical behavior of the persistence probability, however, one can show that an approximate exponential form works well.
 In general $P_\pm (t)$ decays approximately as $\alpha + \beta\exp(-\gamma t )$, with $\gamma$ showing 
 non-universality and dependence on the parameters. For the MS model, results are
 almost $q$ independent. Such $q$ independence had been observed for 
 the steady state behavior earlier but it does not follow that 
 persistence probability will also behave similarly.  

The heatmap of the saturation values of the persistence probability 
 for different initial density of positive opinions are presented for the DMSS model. It has a close correspondence 
 with the  equilibrium phase boundaries obtained on the basis of consensus formation, the agreement becomes better with higher values of $c_0$.  
 Hence, although, in general, persistence probability is not determined by the equilibrium properties of a system, here we find an interesting exception.
 


 In the DMSS model, we find that persistence probability can saturate as long as 
 at least one or both of $\varepsilon_\uparrow$ and $\varepsilon_\downarrow$ is smaller than $1/q$. 
 Since these are the flipping probabilities, this is not an unexpected behavior. Also, the heatmaps show that as the flipping probabilities increase, the saturation values tend to decrease.


Although we could not obtain closed form expressions for the persistence probabilities, we note that the leading order behavior is exponential, as manifested by the plots. At each time step, any agent is picked up for updating and there is always a finite probability of changing their state. Hence the persistence probability is not expected to have a power law behavior in contrast to Ising or Potts model. In  case a stable consensus state is approached, at long times, the $q$ panel would be unanimous and naturally one type of opinion will remain unchanged. However, it is still not imperative that the persistence probability corresponding to that opinion will be non-zero, as that is determined by the states throughout the evolution. For example, in the Voter model in two dimensions,
although a consensus state is reached, the persistence probability decays in a stretched exponential manner and does not attain a saturation value. Hence the results indicate a  unique and non-trivial behavior of the persistence probability when compared to Ising and voter like models.

In  the DMSS model,  when there is no stable consensus state, still  a majority can be reached when $c \neq 1/2$, i.e., we have an excess of either positive or negative opinion overall. 
This may be regarded as closer to reality as it is highly unlikely that all the votes are cast in favour of one candidate (in case of elections) or 
for/against the motion (in case of referendums). In such situations, we find that the persistence goes down exponentially in time, irrespective of the type of persistence. Our results indicate that
when opinions are formed purely from interactions with others, they are bound to change within finite times in realistic situations, i.e., when no consensus is reached. The MS model, on the other hand,
always leads to consensus states except for $p=1/2$, and similarly we find $P_\pm$ attaining a nonzero saturation value for positive/negative consensus. So although the two models are biased in a different way, the qualitative 
behavior of persistence probability is determined by the equilibrium fixed points in both.\\

\begin{acknowledgments}
AP would like to acknowledge University Grant Commission (UGC), Govt. of India, for financial support (NTA REFERENCE NO.: 241610061476). PS acknowledges funding from RUSA 2.0 (Ministry of Education, Government of India) grant.
\end{acknowledgments}

\bibliographystyle{model1-num-names}

\bibliography{refs_pers}

@article{mullick2025social,
  title={Social influence and consensus building: Introducing a q-voter model with weighted influence},
  author={Mullick, Pratik and Sen, Parongama},
  journal={PLoS One},
  volume={20},
  number={1},
  pages={e0316889},
  year={2025},
  publisher={Public Library of Science San Francisco, CA USA}
}

@article{jusup2022social,
  title={Social physics},
  author={Jusup, Marko and Holme, Petter and Kanazawa, Kiyoshi and Takayasu, Misako and Romi{\'c}, Ivan and Wang, Zhen and Ge{\v{c}}ek, Sun{\v{c}}ana and Lipi{\'c}, Tomislav and Podobnik, Boris and Wang, Lin and others},
  journal={Physics Reports},
  volume={948},
  pages={1--148},
  year={2022},
  publisher={Elsevier}
}

@article{castellano2009nonlinear,
  title={Nonlinear q-voter model},
  author={Castellano, Claudio and Mu{\~n}oz, Miguel A and Pastor-Satorras, Romualdo},
  journal={Physical Review E—Statistical, Nonlinear, and Soft Matter Physics},
  volume={80},
  number={4},
  pages={041129},
  year={2009},
  publisher={APS}
}

@book{sen2014sociophysics,
  title={Sociophysics: an introduction},
  author={Sen, Parongama and Chakrabarti, Bikas K},
  year={2014},
  publisher={OUP Oxford}
}

@article{castellano2009statistical,
  title={Statistical physics of social dynamics},
  author={Castellano, Claudio and Fortunato, Santo and Loreto, Vittorio},
  journal={Reviews of modern physics},
  volume={81},
  number={2},
  pages={591--646},
  year={2009},
  publisher={APS}
}

@article{muslim2024mass,
  title={Mass media and its impact on opinion dynamics of the nonlinear q-voter model},
  author={Muslim, Roni and Nqz, Rinto Anugraha and Khalif, Muhammad Ardhi},
  journal={Physica A: Statistical Mechanics and its Applications},
  volume={633},
  pages={129358},
  year={2024},
  publisher={Elsevier}
}

@article{Satya_rev,
author = {Alan J. Bray and Satya N. Majumdar and Grégory Schehr},
title = {Persistence and first-passage properties in nonequilibrium systems},
journal = {Advances in Physics},
volume = {62},
number = {3},
pages = {225--361},
year = {2013},
publisher = {Taylor \& Francis},
doi = {10.1080/00018732.2013.803819},


URL = { 
    
        https://doi.org/10.1080/00018732.2013.803819
    
    

},
eprint = { 
    
        https://doi.org/10.1080/00018732.2013.803819
    
    

}

}

@article{Derrida,
  title = {Exact First-Passage Exponents of 1D Domain Growth: Relation to a Reaction-Diffusion Model},
  author = {Derrida, Bernard and Hakim, Vincent and Pasquier, Vincent},
  journal = {Phys. Rev. Lett.},
  volume = {75},
  issue = {4},
  pages = {751--754},
  numpages = {0},
  year = {1995},
  month = {Jul},
  publisher = {American Physical Society},
  doi = {10.1103/PhysRevLett.75.751},
  url = {https://link.aps.org/doi/10.1103/PhysRevLett.75.751}
}

@article{Stauffer_1994,
doi = {10.1088/0305-4470/27/14/027},
url = {https://doi.org/10.1088/0305-4470/27/14/027},
year = {1994},
month = {jul},
publisher = {},
volume = {27},
number = {14},
pages = {5029},
author = {D Stauffer},
title = {Ising spinodal decomposition at T=O in one to five dimensions},
journal = {Journal of Physics A: Mathematical and General}
}

@article{Surface_dasgupta,
  title = {Persistence in nonequilibrium surface growth},
  author = {Constantin, M. and Dasgupta, C. and Chatraphorn, P. Punyindu and Majumdar, Satya N. and Das Sarma, S.},
  journal = {Phys. Rev. E},
  volume = {69},
  issue = {6},
  pages = {061608},
  numpages = {22},
  year = {2004},
  month = {Jun},
  publisher = {American Physical Society},
  doi = {10.1103/PhysRevE.69.061608},
  url = {https://link.aps.org/doi/10.1103/PhysRevE.69.061608}
}

@article{HILHORST2000124,
title = {Persistence exponent of the diffusion equation in $\varepsilon$ dimensions},
journal = {Physica A: Statistical Mechanics and its Applications},
volume = {277},
number = {1},
pages = {124-126},
year = {2000},
issn = {0378-4371},
doi = {https://doi.org/10.1016/S0378-4371(99)00509-9},
url = {https://www.sciencedirect.com/science/article/pii/S0378437199005099},
author = {H.J Hilhorst}
}

@article{PhysRevLett.77.2867,
  title = {Nontrivial Exponent for Simple Diffusion},
  author = {Majumdar, Satya N. and Sire, Cl\'ement and Bray, Alan J. and Cornell, Stephen J.},
  journal = {Phys. Rev. Lett.},
  volume = {77},
  issue = {14},
  pages = {2867--2870},
  numpages = {0},
  year = {1996},
  month = {Sep},
  publisher = {American Physical Society},
  doi = {10.1103/PhysRevLett.77.2867},
  url = {https://link.aps.org/doi/10.1103/PhysRevLett.77.2867}
}

@article{Surface_Krug,
  title = {Persistence exponents for fluctuating interfaces},
  author = {Krug, J. and Kallabis, H. and Majumdar, S. N. and Cornell, S. J. and Bray, A. J. and Sire, C.},
  journal = {Phys. Rev. E},
  volume = {56},
  issue = {3},
  pages = {2702--2712},
  numpages = {0},
  year = {1997},
  month = {Sep},
  publisher = {American Physical Society},
  doi = {10.1103/PhysRevE.56.2702},
  url = {https://link.aps.org/doi/10.1103/PhysRevE.56.2702}
}

@article{Brownian_bennaim,
  title = {Slow Kinetics of Brownian Maxima},
  author = {Ben-Naim, E. and Krapivsky, P. L.},
  journal = {Phys. Rev. Lett.},
  volume = {113},
  issue = {3},
  pages = {030604},
  numpages = {5},
  year = {2014},
  month = {Jul},
  publisher = {American Physical Society},
  doi = {10.1103/PhysRevLett.113.030604},
  url = {https://link.aps.org/doi/10.1103/PhysRevLett.113.030604}
}

@article{Surface_sakagawa,
 ISSN = {00018678},
 URL = {http://www.jstor.org/stable/43563465},
 author = {HIRONOBU SAKAGAWA},
 journal = {Advances in Applied Probability},
 number = {1},
 pages = {146--163},
 publisher = {Applied Probability Trust},
 title = {PERSISTENCE PROBABILITY FOR A CLASS OF GAUSSIAN PROCESSES RELATED TO RANDOM INTERFACE MODELS},
 urldate = {2026-02-16},
 volume = {47},
 year = {2015}
}

@article{voter_pers,
  title = {Coarsening and persistence in the voter model},
  author = {Ben-Naim, E. and Frachebourg, L. and Krapivsky, P. L.},
  journal = {Phys. Rev. E},
  volume = {53},
  issue = {4},
  pages = {3078--3087},
  numpages = {0},
  year = {1996},
  month = {Apr},
  publisher = {American Physical Society},
  doi = {10.1103/PhysRevE.53.3078},
  url = {https://link.aps.org/doi/10.1103/PhysRevE.53.3078}
}

@article{Sznajd_pers,
    author = {Stauffer, D and de Oliveira, P.M.C.},
    title = {Persistence of opinion in the Sznajd consensus model: computer simulation},
    journal = {The European Physical Journal B - Condensed Matter and Complex Systems},
    year = {2002},
     volume = {30},
    pages = {587--592}
}

@article{Soham2009,
  title = {Model of binary opinion dynamics: Coarsening and effect of disorder},
  author = {Biswas, Soham and Sen, Parongama},
  journal = {Phys. Rev. E},
  volume = {80},
  issue = {2},
  pages = {027101},
  numpages = {4},
  year = {2009},
  month = {Aug},
  publisher = {American Physical Society},
  doi = {10.1103/PhysRevE.80.027101},
  url = {https://link.aps.org/doi/10.1103/PhysRevE.80.027101}
}

@article{Sudip2020,
  title = {Long route to consensus: Two-stage coarsening in a binary choice voting model},
  author = {Mukherjee, Sudip and Biswas, Soumyajyoti and Sen, Parongama},
  journal = {Phys. Rev. E},
  volume = {102},
  issue = {1},
  pages = {012316},
  numpages = {9},
  year = {2020},
  month = {Jul},
  publisher = {American Physical Society},
  doi = {10.1103/PhysRevE.102.012316},
  url = {https://link.aps.org/doi/10.1103/PhysRevE.102.012316}
}

@article{gen_qvoter,
    author = {Doniec, Maciej and Mullick, Pratik and Sen, Parongama and Sznajd-Weron, Katarzyna},
    title = {Modeling biases in binary decision-making within the generalized nonlinear q-voter model},
    journal = {Chaos: An Interdisciplinary Journal of Nonlinear Science},
    volume = {35},
    number = {4},
    pages = {043133},
    year = {2025},
    month = {04},
    issn = {1054-1500},
    doi = {10.1063/5.0266510},
    url = {https://doi.org/10.1063/5.0266510},
    eprint = {https://pubs.aip.org/aip/cha/article-pdf/doi/10.1063/5.0266510/20490567/043133_1_5.0266510.pdf},
}

@article{appmps2026,
  title={Analyzing contrarian behavior using nonlinear biased q-voter model},
  author={Pradhan, Amit and Mullick, Pratik and Sen, Parongama},
  journal={Physical Review E},
  volume={113},
  number={3},
  pages={034301},
  year={2026},
  publisher={APS}
}

@article{biswas2017critical,
  title={Critical noise can make the minority candidate win: The US presidential election cases},
  author={Biswas, Soumyajyoti and Sen, Parongama},
  journal={Physical Review E},
  volume={96},
  number={3},
  pages={032303},
  year={2017},
  publisher={APS}
}

\end{document}